\documentclass[conference]{IEEEtran}
\IEEEoverridecommandlockouts
\usepackage{cite}
\usepackage{amsmath,amssymb,amsfonts}
\usepackage{algorithmic}
\usepackage{graphicx}
\usepackage{textcomp}
\usepackage{xcolor}
\def\BibTeX{{\rm B\kern-.05em{\sc i\kern-.025em b}\kern-.08em
    T\kern-.1667em\lower.7ex\hbox{E}\kern-.125emX}}

\usepackage{cite}
\usepackage{amsmath,amssymb,amsfonts}
\usepackage{algorithmic}
\usepackage{graphicx}
\usepackage{textcomp}
\usepackage{xcolor}
\usepackage{multicol, blindtext}
\usepackage{xspace,empheq,fancybox,amsmath,amssymb,graphicx,epstopdf,epsfig,syntonly,times,amsthm} \usepackage{psfrag,color,bm,array,cite}
\usepackage{url,cite,footnote,xspace,syntonly,algorithm,algorithmic}
\usepackage{verbatim,multirow}
\usepackage[T1]{fontenc}
\usepackage{mathtools}
\usepackage{amsmath,amsfonts,amsthm,bm}
\usepackage{graphicx}
\usepackage{mwe}
\usepackage{caption}
\usepackage{stfloats}
\usepackage{amsfonts}
\usepackage{stackengine}
\usepackage{lipsum}
\usepackage{subfig}
\usepackage{bm}
\usepackage{booktabs}
\usepackage{tabularx}
\usepackage{bbm}
\usepackage{dsfont}

\def\BibTeX{{\rm B\kern-.05em{\sc i\kern-.025em b}\kern-.08em
    T\kern-.1667em\lower.7ex\hbox{E}\kern-.125emX}}
    
\newtheorem{remark}{\bfseries Remark}

\newcommand\barbelow[1]{\stackunder[1.2pt]{$#1$}{\rule{.9ex}{.115ex}}}
\newcommand{\overbar}[1]{\mkern 0.8mu\overline{\mkern-1mu#1\mkern-1mu}\mkern 0.8mu}

\newcolumntype{P}[1]{>{\centering\arraybackslash}p{#1}}

\newtheorem{assumption}{\hspace{0pt}\bf Assumption}

\begin{document}

\author{
\IEEEauthorblockN{Yuqi Zhou$^\dag$, Cong Feng$^\dag$, Mingzhi Zhang$^\ddag$, and Rui Yang$^\dag$}
\IEEEauthorblockA{$^\dag$Power Systems Engineering Center, $^\ddag$Center for Integrated Mobility Sciences\\National Renewable Energy Laboratory\\Golden, Colorado, USA\\
}
}

\title{Exploring the Use of Autonomous Unmanned Vehicles for Supporting Power Grid Operations\\
\thanks{This work was authored by the National Renewable Energy Laboratory, operated by Alliance for Sustainable Energy, LLC, for the U.S. Department of Energy (DOE). 
Funding provided by U.S. Department of Energy Office of Energy Efficiency and Renewable Energy Solar Energy Technologies Office (SETO) under award DE-EE0052743.
The views expressed in the article do not necessarily represent the views of the DOE or the U.S. Government. The U.S. Government retains and the publisher, by accepting the article for publication, acknowledges that the U.S. Government retains a nonexclusive, paid-up, irrevocable, worldwide license to publish or reproduce the published form of this work, or allow others to do so, for U.S. Government purposes.
}
}

\IEEEaftertitletext{\vspace{-1.05\baselineskip}}

\maketitle

\begin{abstract}

This paper explores the use of autonomous unmanned vehicles to support power grid operations. With built-in batteries and the capability to carry additional battery energy storage, the rising number of autonomous vehicles can represent a substantial amount of capacity that is currently underutilized in the power grid. Unlike traditional electric vehicles that require drivers, the operations of autonomous vehicles can be performed without human intervention. To guide idle vehicles to autonomously support power grids, we propose a tractable optimization-based method to effectively integrate these ``mobile batteries'' into grid operations. During real-time operations, the vehicles are strategically routed to target locations to maintain power balance and reduce operating costs. Numerical studies have confirmed both the validity and the scalability of the proposed algorithm to efficiently integrate autonomous vehicles into routine power system operations.
\end{abstract}

\begin{IEEEkeywords}
Unmanned vehicles, power systems, transportation systems, optimal power flow, mixed-integer optimization
\end{IEEEkeywords}

\vspace{-1mm}

\section{Introduction}

The rapid advancements in robotics, control theory, and artificial intelligence in recent years have positioned us closer than ever to the imminent era of autonomous unmanned vehicles (also known as self-driving or autonomous vehicles). Autonomous vehicles are uncrewed vehicles that are capable of independent operations without human intervention. These vehicles use cutting-edge sensors and complex algorithms to navigate, make real-time decisions, and even interact with surroundings. The gradual expansion of autonomous vehicles is underway in many places across the United States and around the world, with expectations for further growth in the coming years. For example, Waymo has recently launched its self-driving robotaxi service for the general public in Los Angeles \cite{waymo}. 
In addition, there are plenty of autonomous vehicles under development, such as Tesla Cybercab and self-driving trucks (from Kodiak, Aurora, etc). 
One significant yet unexplored use of autonomous vehicles is utilizing their mobility and flexibility to provide essential support to power system operations. During idle time, these vehicles can be dispatched to target locations to help maintain power balance, provide frequency response, and support emergency operations. Given the increasing strain on the U.S. power grid in recent years, these autonomous vehicles can serve as flexible resources to effectively alleviate grid stress. Addressing this research problem could save billions of dollars by reducing power outages and enhancing grid resilience without the necessity of major changes to the power grid infrastructure.

Although extensive research has examined the integration of electric vehicles (EVs) into power systems (e.g., \cite{lopes2010integration,ota2011autonomous,taghizadegan2022dominated}), research on using autonomous vehicles to support grid operations has been very limited. Compared with traditional EVs, autonomous vehicles bring key advantages to power grid support, including \textbf{enhanced operational autonomy}, \textbf{higher flexibility}, and \textbf{improved safety} under challenging weather and road conditions.
Some studies \cite{kim2018enhancing,saboori2021optimal} have explored the role of mobile energy storage units in grid operations, often in the form of trailer-mounted batteries that require separate transport. However, the specifications for the energy storage units such as storage capacity, modeling, and availability, differ from those of autonomous vehicles. Other research has examined autonomous vehicles for other applications \cite{zhou2020design,paparella2024electric}, but the integration into power system operations remains underexplored.
In this work, we aim to develop a framework for integrating autonomous vehicles into grid operations. The task involves decision-making that balances the needs of power systems and the constraints of vehicles and transportation systems. Hence, we first present an optimal routing formulation, which enables each autonomous vehicle to quickly determine the best routes and costs to candidate locations. After that, we introduce a mixed-integer optimization model that integrates autonomous vehicles to assist power systems in maintaining power balance while minimizing the total operating cost. To tackle the nonlinearity in the optimization problem, we leverage McCormick relaxation to develop an exact and tractable reformulation, which enables real-time routing and dispatch of autonomous vehicles to support routine power grid operations.

This paper is organized as follows. Section \ref{sec:routing} introduces the optimal routing formulation for autonomous vehicles in a transportation network. Section \ref{sec:optimization} presents a tractable optimization formulation for integrating autonomous vehicles into grid operations. Section \ref{sec:ns} uses the IEEE 14-bus system for the numerical simulations, with additional systems tested to evaluate the scalability of the algorithm. Section \ref{sec:con} concludes the paper and outlines directions for future research.

\section{Optimal Routing Formulation}
\label{sec:routing}

\begin{figure}[t!]
\centering
\includegraphics[trim=0cm 0cm 0cm 0cm,clip=true,totalheight=0.12\textheight]{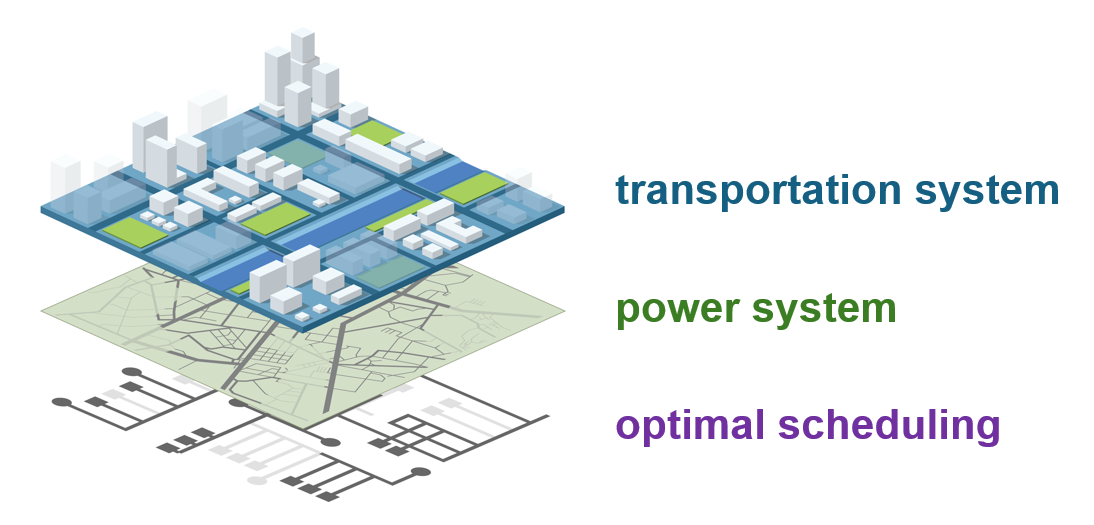}
\caption{A demonstration of integrated decision-making under both transportation and power system layers.}
\label{fig:demonstration}
\vspace{-3mm}
\end{figure}

The incorporation of autonomous vehicles to support grid operations poses an integrated decision-making problem under two layers (see Fig. \ref{fig:demonstration}): One is the transportation system that the vehicles operate on, and the other is the power system that the vehicles intend to interact with. First, we focus on the transportation layer, and we introduce an optimization formulation for the autonomous vehicle routing problem, assuming that the starting point and destination have been determined. Consider a transportation network represented by a directed graph, $G$, with a total of $N$ nodes collected in the set $\cal N :=$ $\{1,\ldots,N\}$. To simplify the modeling, we assume that the transportation network shares the same set of nodes as the power network, with different connections among them, which is a widely used assumption in mobile energy storage integration problems (see, e.g., 
\cite{kim2018enhancing,dugan2021application}). The branches of the transportation network are collected in the set $\cal E :=$ $\{(i,j)\} \subset \cal N \times \cal N$, where the weight of each branch is represented by $w_{ij}$. In the vehicle routing problem, these weights are typically positive, and they represent known features associated with each branch, such as distance, congestion, or tolls. 

The goal of the optimal routing algorithm is to determine the most cost-effective route for an autonomous vehicle between an origin point, $s$, and a destination point, $e$. To this end, we can define the binary decision variables $x_{ij} \in \{0,1\}$ on each branch, where $x_{ij} = 1$ indicates that the branch is included in the optimal route, and $x_{ij} = 0$ means it is not part of the route. With this information, we can enforce several network constraints to help identify the optimal route. To begin, the route must initiate from the starting node $s$; thus:
\begin{align}
    \sum_{j \in {\cal{A}}_{s}} x_{sj} = 1
\label{eq:con_1}
\end{align}
where ${\cal{A}}_{s}$ denotes the set of neighboring nodes of node $s$. On the other hand, the vehicle must also stop at the destination point, $e$, leading to the following:
\begin{align}
    \sum_{j \in {\cal{A}}_{e}} x_{ej} = 0 
\label{eq:con_2}
\end{align}
For all nodes other than the origin or the destination, the flow conservation imposes additional network constraints:
\begin{align}
    \sum_{i \in {\cal{A}}_{k}} x_{i, k} = \sum_{j \in {\cal{A}}_{k}} x_{k, j}, \ \forall k \in {\cal N}, \, k \neq s, \, k \neq e.
\label{eq:con_3}
\end{align}
In addition, to guarantee that the optimal route passes through each branch no more than once, the following linear constraint is enforced on each pair of decision variables to ensure the paths do not overlap during the search for the routing decision:
\begin{align}
    x_{ij} + x_{ji} \leq 1, \quad \forall (i,j) \in \cal{E}
\label{eq:con_4}
\end{align}
Furthermore, for certain paths with restrictions (e.g., limited-access roads, road closures), the corresponding variables are enforced to be zero:
\begin{align}
    x_{ij} = 0, \quad \forall (i,j) \in \cal{E'} \subset \cal{E}
\label{eq:con_5}
\end{align}
where $\cal{E'}$ denotes the set of restricted paths.

Based on these network constraints, the optimal routing problem can be formulated as:
\begin{subequations}
\label{eq:routing}
\begin{align}
\min \quad & \sum_{(i,j) \in \cal{E}} w_{ij} x_{ij} \\
\textrm{s.t.} \quad \: 
  & x_{ij} \in \{0,1\}, \quad \forall (i,j) \in \cal{E}\\
  & \text{Eqs. } \eqref{eq:con_1}-\eqref{eq:con_5}
\end{align}
\end{subequations}
which is a mixed-integer linear program that can be efficiently solved using optimization solvers, such as CPLEX and Gurobi. Note that the objective function in this formulation considers a comprehensive weight for each branch, but it can be easily modified to accommodate multiple objectives.

\begin{remark}[Choice of routing algorithm]
Although numerous algorithms including Dijkstra's algorithm, Bellman-Ford algorithm, and $A^{*}$ algorithm, are available for solving the routing problem in graphs, their performance varies depending on the network structure and problem constraints. In this work, we formulate it as a generalized optimization problem because it allows for the incorporation of specific constraints and uncertain parameters (e.g., stochastic/robust optimization) for further development of the algorithm.
\end{remark}

\begin{figure}[t!]
\centering
\includegraphics[trim=1cm 0cm 0cm 0cm,clip=true,totalheight=0.12\textheight]{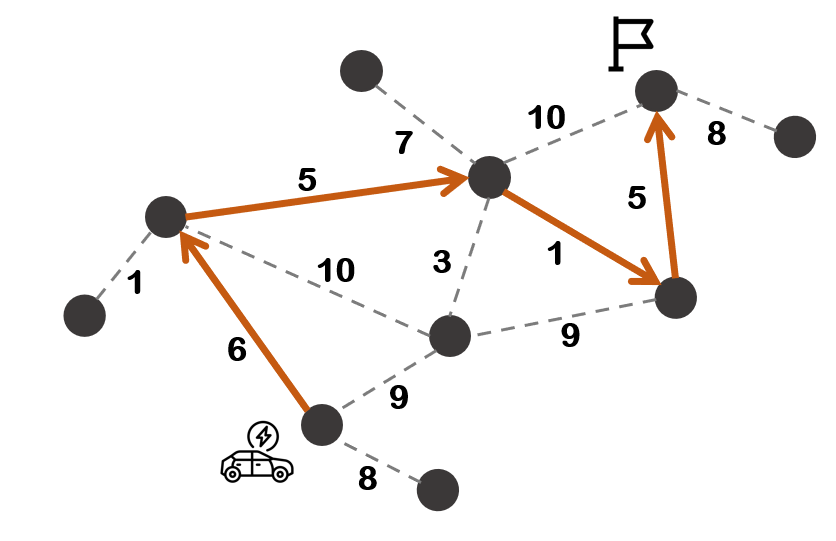}
\caption{The optimal route of a vehicle in a 10-node network after solving the problem \eqref{eq:routing}, with an optimal value of 17.}
\label{fig:routing_example}
\vspace{-2mm}
\end{figure}

Once the optimization problem \eqref{eq:routing} is solved, the total travel cost for the vehicle under consideration can be expressed as:
\begin{align}
    c^{v} = \sum_{(i,j) \in \cal{E}} w_{ij} x^{*}_{ij} = w_{s{k_1}} + w_{{k_1}{k_2}} + \cdots + w_{{k_n}e}
\label{eq:routing_cost}
\end{align}
in which $x^{*}$ denotes the optimal solution of $x$. Further, the optimal route can be represented using the set:
\begin{align}
    {\cal{R}}^{v} = \{(i,j) \mid x^{*}_{ij} = 1\}
\label{eq:routing_set}
\end{align}
which consists of all intermediate paths with nonzero $x$ values from the origin to the destination.

\section{Integrating and Optimizing Autonomous Unmanned Vehicles for Grid Operations}
\label{sec:optimization}

After completing the transportation modeling, we will shift our focus to power grid modeling and the integration of autonomous vehicles to support grid operations. To this end, let us consider a power grid with all buses collected in the set $\cal N :=$ $\{1,\ldots,N\}$ and all lines collected in the set $\cal L$. For each bus $i$, denote $\theta_i$ as its phase angle, $p^{g}_{i}$ and $p^{d}_{i}$ as its connected generation and load, respectively. The phase angle for the reference bus is set to $0$, and thus $\theta_{\text{ref}} = 0$. For each line $(i,j)$, its line flow is denoted by $f_{ij}$. Under the dc power flow model \cite{stott2009dc}, the line flow from bus $i$ to bus $j$ is given by:
\begin{align}
    f_{ij} = b_{ij}(\theta_i - \theta_j), \: \: \forall (i,j) \in \cal{L}
\end{align}
where $b_{ij}$ represents the inverse of line reactance.

In modern power systems, day-ahead and intra-day optimal power flow calculations are becoming more common. System operators can adjust their schedules and dispatch plans (e.g., minutes to hours before real time) to accommodate changes in demand, renewable generation, and other distributed energy resources. Note that autonomous vehicles need time to travel; for simplicity, we assume that they can safely arrive at target locations within the planned operational time frame after being dispatched. In addition, regarding the power delivered to the grid by different autonomous vehicles, we make the following assumption:
\begin{assumption}[Power generation from autonomous vehicles]
The integration of autonomous vehicles to support grid operations is mainly achieved through technologies such as V2G \cite{karfopoulos2015distributed}. For ease of modeling, in this work, we ignore the state of charge and discharging efficiency of batteries, and simply provide a nominal power generation value $p^{v}$ for each vehicle $v$. We assume that all the selected vehicles $v \in \cal{V}$ can provide power generation reliably during the limited time frame, once dispatched to their designated locations.
\end{assumption}
The integration of autonomous vehicles for grid operations can be formulated as an optimal power flow-based problem, as given by:
\begin{subequations}
\label{eq:opf}
\begin{align}
\min \quad & \sum_{i \in \cal{G}} c^{g}_{i} (p^{g}_{i}) + \sum_{v \in \cal{V}} c^{v} (p^{v}) + \sum_{v \in \cal{V}} \sum_{i \in \cal{N}} c^{v}_{i} z^{v}_{i}  \label{eq:o_a}\\
\textrm{s.t.} \quad \: 
  & \barbelow{p}^{g}_{i} \leq p^{g}_{i} \leq \overbar{p}^{g}_{i}, \quad \forall i \label{eq:o_b}\\
  & {\barbelow{p}}^{v} \leq p^{v} \leq {\overbar{p}}^{v}, \quad \forall v \label{eq:o_c}\\
  & \barbelow{f}_{ij} \leq b_{ij}(\theta_{i} - \theta_{j}) \leq \overbar{f}_{ij}, \quad \forall (i,j) \label{eq:o_d}\\
  & z^{v}_{i} \in \{0,1\}, \quad \forall i, v \label{eq:o_e}\\
  & 0 \leq \sum_{i \in {\cal{N}}} z^{v}_{i} \leq 1, \quad \forall v \label{eq:o_f}\\
  & \sum_{j \in {\cal{A}}_{i}} b_{ij} (\theta_i - \theta_j) = p^{g}_{i} - p^{d}_{i} + \sum_{v \in \cal{V}} p^{v} z^{v}_{i}, \: \forall i \label{eq:o_g}
\end{align}
\end{subequations}
The decision variables include both the power from traditional generators $p^{g}$, and the power provided by autonomous vehicles $p^{v}$. Furthermore, the problem determines the line flow $f_{ij}$ for each line as well as the phase angle $\theta_{i}$ per bus. Last, a binary decision variable $z^{v}_{i}$ is introduced to indicate whether a vehicle $v$ travels from its origin to node $i$ ($z^{v}_{i} = 1$: positive, $z^{v}_{i} = 0$: negative). The objective function \eqref{eq:o_a} consists of generation costs from traditional generators $c^{g}_{i} (\cdot)$ and autonomous vehicles $c^{v} (\cdot)$, which are typically (piecewise) linear functions. In addition, the objective function also includes the total travel cost such as electricity costs, battery degradation, and toll fees, of each vehicle (see equation \eqref{eq:routing_cost}), given by $\sum_{i \in \cal{N}} c^{v}_{i} z^{v}_{i}$. As for constraints, \eqref{eq:o_b} and \eqref{eq:o_c} provide the generation limits for generators and vehicles. Line flow constraints are given in \eqref{eq:o_d}. Constraints \eqref{eq:o_e} and \eqref{eq:o_f} ensure that each vehicle can travel to at most one location. The power balance for each bus is enforced in \eqref{eq:o_g}. For large-scale systems, long-distance vehicle travel might be unrealistic to effectively provide grid support. Therefore, the following constraint can also be added to exclude nodes that are too far from the vehicle:
\begin{align}
    z^{v}_{i} = 0, \quad \forall i \in {\cal{N}}_{v} \subseteq {\cal{N}}, \: \forall v \label{eq:far}
\end{align}
Note that \eqref{eq:opf} is a nonlinear optimization problem at this point, due to the bilinear product $p^{v} z^{v}_{i}$ in constraint \eqref{eq:o_g}. As a result, solving it before the dispatch of autonomous vehicles can be computationally challenging. To address this issue, we use the McCormick relaxation \cite{gupte2013solving} technique to reformulate the problem, making it suitable for standard mixed-integer optimization solvers. To that end, we first define the product of $p^{v}$ and $z^{v}_{i}$, $\forall i \in \cal{N}$, $\forall v \in \cal{V}$ as the following:
\begin{align}
    y^{v}_{i} = p^{v} z^{v}_{i} \label{eq:product}
\end{align}
With lower/upper bounds ${\barbelow{p}}^{v}, {\overbar{p}}^{v}$ given for the variable $p^{v}$, the following linear inequalities hold:
\begin{subequations} \label{eq:mc}
\begin{align}
& y^{v}_{i} \geq z^{v}_{i} {\barbelow{p}}^{v},\label{eq:mc_a}\\
& y^{v}_{i} \geq p^{v} + z^{v}_{i} {\overbar{p}}^{v} - {\overbar{p}}^{v},\label{eq:mc_b}\\
& y^{v}_{i} \leq z^{v}_{i} {\overbar{p}}^{v},\label{eq:mc_c}\\
& y^{v}_{i} \leq p^{v} + z^{v}_{i} {\barbelow{p}}^{v} - {\barbelow{p}}^{v}. \label{eq:mc_d}
\end{align}
\end{subequations}

On one hand, the inequalities \eqref{eq:mc_a}-\eqref{eq:mc_d} can be easily verified by substituting \eqref{eq:product}. On the other hand, these constraints also jointly guarantee the equality \eqref{eq:product} under bilinear multiplication. Specifically, if $z^{v}_{i}$ equals zero, constraints \eqref{eq:mc_a} and \eqref{eq:mc_c} together enforce $y^{v}_{i} = 0$. If $z^{v}_{i}$ is equal to one, constraints \eqref{eq:mc_b} and \eqref{eq:mc_d} together guarantee that $y^{v}_{i} = p^{v}$. Therefore, the inequalities in \eqref{eq:mc} are \textit{equivalent} to the bilinear equation \eqref{eq:product}, and the nonlinear optimization \eqref{eq:opf} can be transformed into the following tractable formulation:
\begin{subequations}
\label{eq:tractable}
\begin{align}
\min \quad & \sum_{i \in \cal{G}} c^{g}_{i} (p^{g}_{i}) + \sum_{v \in \cal{V}} c^{v} (p^{v}) + \sum_{v \in \cal{V}} \sum_{i \in \cal{N}} c^{v}_{i} z^{v}_{i} \label{eq:tractable_a}\\
\textrm{s.t.} \quad \: 
  & \text{Eqs. } \eqref{eq:o_b}-\eqref{eq:o_f}, \eqref{eq:mc_a}-\eqref{eq:mc_d}\\
  & \sum_{j \in {\cal{A}}_{i}} b_{ij} (\theta_i - \theta_j) = p^{g}_{i} - p^{d}_{i} + \sum_{v \in \cal{V}} y^{v}_{i}, \: \forall i
\end{align}
\end{subequations}
Because the bilinear term in \eqref{eq:product} involves at least one binary variable, the convex relaxation in \eqref{eq:mc} is exact, and this optimization problem is an \textit{exact reformulation} of the problem \eqref{eq:opf}. The reformulation using the linear inequalities in \eqref{eq:mc} is known as the McCormick relaxation \cite{gupte2013solving}, and it has been widely used in other topology design and scheduling problems \cite{singh2019optimal,zhou2021efficient,lv2022security}. Instead of handling the original nonlinear program, we now have a mixed-integer program that is efficiently solvable. By using the optimal solutions of $z^{v}_{i}$, we can easily determine the assigned destination for each vehicle:
\begin{align} \label{eq:mapping}
    e^{v} = \underset{i \in \mathcal{N}}{\text{argmax}} \: z^{v*}_i , \quad \forall v 
    \in \cal{V}
\end{align}
Once the destination is known, the optimal route for each vehicle can be obtained from the solutions to problem \eqref{eq:routing}. The detailed process of integrating autonomous vehicles into grid operations is provided in Algorithm \ref{alg:alg1}. To summarize, the optimal routing problem \eqref{eq:routing} is first solved repeatedly offline to determine the travel costs. Afterward, the optimization problem \eqref{eq:tractable} is solved to determine each vehicle's travel destination for grid support. Notably, as each vehicle travels, the optimal routing problem can be re-solved in real time to account for changes in road conditions and adjust the route accordingly.

\begin{figure}[!t] 
  \begin{algorithm}[H]
 \caption{Optimal Scheduling of Autonomous Vehicles}
 \label{alg:alg1}
 \begin{algorithmic}[1]
 \renewcommand{\algorithmicrequire}{\textbf{Inputs:}}
 \renewcommand{\algorithmicensure}{\textbf{Outputs:}}
 \REQUIRE $w_{ij}, c^{g}_{i} (\cdot), c^{v} (\cdot), p^{d}_{i}, \textit{topology}, \textit{operational limits}$
 \ENSURE  $e^{v}, {\cal{R}}^{v}, \ v \in \cal{V}$
 \\ \textit{Initialization}: \text{Define the set $\cal{E'} \subset \cal{E}$ by road conditions.} 
  \STATE Solve optimization problem \eqref{eq:routing} under different scenarios to obtain the travel cost, $c^{v}$, offline.
  \STATE Determine the set of candidate vehicles, ${\cal{V}}$, to provide grid support.
  \FOR {$v = 1$ to $|{\cal{V}}|$}
  \STATE Confirm the power generation limits, ${\barbelow{p}}^{v}$ and ${\overbar{p}}^{v}$, from each candidate vehicle.    
  \STATE Determine the set of candidate nodes for each vehicle to travel to.
  Set $z^{v}_{i} = 0$ for other nodes ${\cal{N}}_{v} \subseteq {\cal{N}}$.
  \ENDFOR
  \STATE Solve the mixed-integer program \eqref{eq:tractable} and obtain $z^{v*}_{i}$
  \STATE Record the optimal location for each vehicle using \eqref{eq:mapping}.
  \STATE Record the optimal route for each vehicle via \eqref{eq:routing_set}.
 \RETURN $e^{v}, {\cal{R}}^{v}, \ v \in \cal{V}$
 \end{algorithmic}
 \end{algorithm}
 \vspace{-20pt} 
\end{figure}

\section{Numerical Simulation}
\label{sec:ns}

\begin{figure}[t!]
\centering
\includegraphics[trim=0cm 0cm 0cm 0cm,clip=true,totalheight=0.16\textheight]{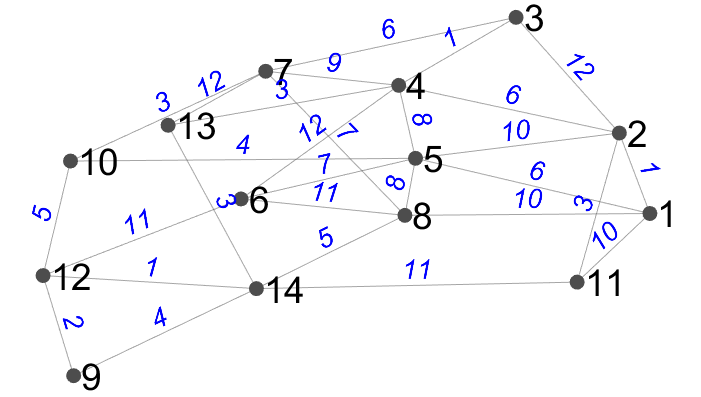}
\caption{A synthetic 14-node meshed transportation network.}
\label{fig:transportation}
\vspace{-4mm}
\end{figure}

In this section, we present the numerical simulation results of the proposed algorithm. The optimization problems are implemented in MATLAB and solved using the Gurobi solver, utilizing 12 available processor threads. The simulations are performed on a regular laptop with an Intel CPU @ 2.60 GHz and 16 GB of RAM. 
We mainly use the IEEE 14-bus system to show the numerical performance, with additional test cases to demonstrate the scalability and computational efficiency of the proposed algorithm. Accordingly, the 14-bus test case, which includes 20 transmission lines and 5 conventional generators, is applied to model the power systems. For the transportation model, we construct a synthetic 14-node interconnected network that overlays with the power system, with weights randomly assigned between 1 and 12. The use of random weights makes the models more versatile, while the costs are designed such that neither generation nor travel costs dominates the objective terms for vehicles in \eqref{eq:tractable_a}.
To better display the connections and weights (in blue), the transportation network topology is slightly reorganized, as shown in Fig. \ref{fig:transportation}. A linear generation cost function is used for both the generators and the autonomous vehicles, each with different cost coefficients. For the numerical simulation, we assume the system has three groups of autonomous vehicles (e.g., vehicle fleets) ready for dispatch. This ensures that the total generation capacity of the autonomous vehicles is comparable to that of traditional generators. For simplicity, we refer to these groups as ``vehicles,'' and thus ${\cal{V}} = \{a,b,c\}$. At the time of consideration, these vehicles are located at buses 2, 7, and 12, and they can be navigated to provide power generation with enough battery state of charge.

\begin{figure}[t!]
\centering
\includegraphics[trim=0cm 0cm 0cm 0cm,clip=true,totalheight=0.18\textheight]{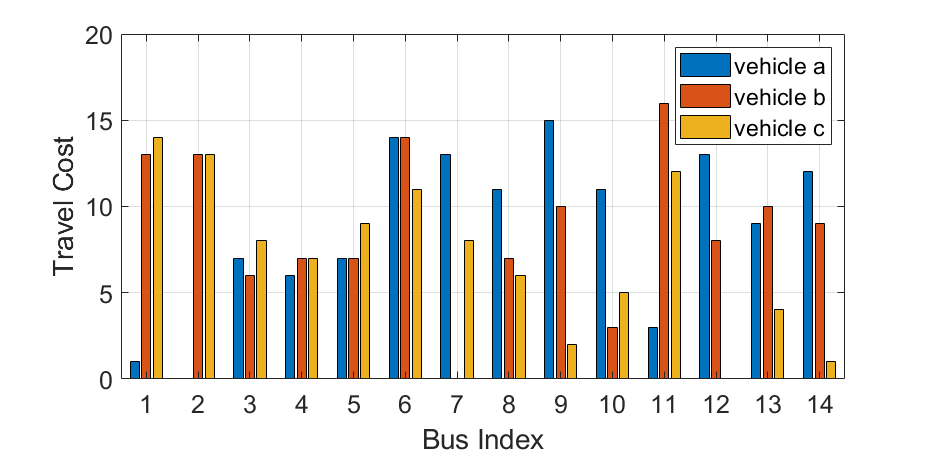}
\caption{The minimum total travel costs of vehicles $v \in \{a,b,c\}$ to every node in the system.}
\label{fig:cost}
\vspace{-4mm}
\end{figure}

\begin{figure}[t!]
\centering
\includegraphics[trim=0.2cm 0cm 0cm 0cm,clip=true,totalheight=0.12\textheight]{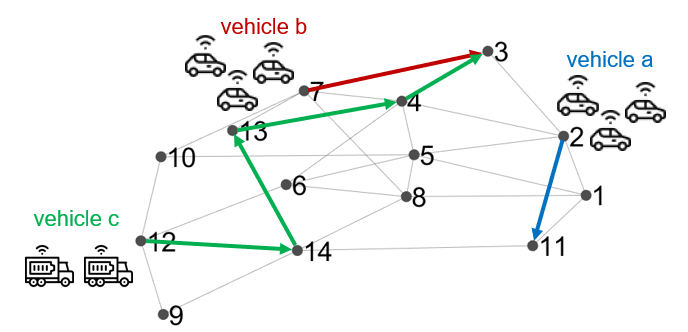}
\caption{Optimal routes for candidate autonomous vehicles.}
\label{fig:simulation}
\vspace{-3mm}
\end{figure}

With all the provided system information, we will now walk through every step of Algorithm \ref{alg:alg1} to demonstrate how to obtain the optimal scheduling of the autonomous vehicles. For the given transportation network shown in Fig.~\ref{fig:transportation}, we assume that all the roads are accessible. Then, we solve the problem \eqref{eq:routing} for the candidate vehicles, $v \in {\cal{V}}$, to find the travel cost to each node in the system. The travel costs for each vehicle to every node in the system are shown in Fig.~\ref{fig:cost}. Under the weights assigned to the transportation network, the total travel cost to each node ranges from 0 to 16, with 0 representing no movement as the vehicle remains at its original point. Note that this routing problem is solved offline, and the minimum travel costs to all nodes will be used as the cost coefficients for $z^{v}_{i}$ in the objective function of the optimization problem \eqref{eq:tractable} that will be solved later. For the autonomous vehicle integration problem \eqref{eq:tractable}, we set the generation limits for each vehicle fleet as ${\overbar{p}}^{a} = 20 \ \text{MW}$, ${\overbar{p}}^{b} = 20 \ \text{MW}$, and ${\overbar{p}}^{c} = 40 \ \text{MW}$, where the last group includes autonomous trucks that can be equipped with additional battery energy storage units to allow for higher generation capacity. The generation capacity provided by these autonomous vehicles accounts for approximately $10.4\%$ of the total capacity of generators. Given the system size, we designate all nodes in the system as candidate nodes for each vehicle. The mixed-integer problem \eqref{eq:tractable} is solved next, and we can obtain the optimal destination/route for each vehicle. The results are demonstrated in Fig.~\ref{fig:simulation}, where vehicle $a$ is routed to node 11, and vehicles $b,c$ are navigated to node 3 to provide generation. Furthermore, the optimal generation by these vehicles is $p_{a}^{*} = 17.28 \ \text{MW}$, $p_{b}^{*} = 20 \ \text{MW}$, $p_{c}^{*} = 40 \ \text{MW}$, where $b, c$ are fully used, and $a$ is the marginal unit. With the integration of autonomous vehicles, the total generation of traditional generators becomes $181.72 \ \text{MW}$, and the total cost is reduced by $26.27\%$. This is largely because the proposed optimization algorithm will strategically navigate these ``mobile batteries'' to the target locations, hence avoiding the use of costly traditional generators.

\begin{table}[t!]
\renewcommand{\arraystretch}{1.1}
\caption{Computation time (in seconds) of the autonomous vehicle integration problem under various test cases.}
    \centering
    \begin{tabular}{|p{2.8cm}|p{1.1cm}|p{1.1cm}|p{1.1cm}|}
        \hline
        \textbf{Test Cases} & \textbf{Min} & \textbf{Median} & \textbf{Max} \\
        \hline
        IEEE 14-bus & 0.4906 & 0.5081 & 0.5747 \\
        \hline
        IEEE 30-bus & 0.5836 & 0.6051 & 0.6205 \\
        \hline
        IEEE 118-bus & 0.7355 & 0.7463 & 0.7757 \\
        \hline
        IEEE 300-bus & 1.1853 & 1.2187 & 1.7006 \\
        \hline
        South Carolina 500-bus & 1.6558 & 1.7822 & 2.0464 \\
        \hline
        European 1354-bus & 6.9108 & 7.1728 & 8.3905 \\
        \hline
    \end{tabular}
    \label{tab:statistics}
    \vspace{-3mm}
\end{table}

To further verify the scalability and computational efficiency of the proposed algorithm, we also test it under various test cases. These systems include IEEE 14-bus, 30-bus, 118-bus, 300-bus, as well as a 500-bus South Carolina synthetic system and a 1354-bus portion of a European transmission system. The computation time for solving the optimization problem \eqref{eq:tractable} under these test cases are recorded in Table~\ref{tab:statistics}. Thanks to the McCormick relaxation-based reformulation, the corresponding optimization problem becomes more tractable to solve. Throughout all the numerical simulations, the runtime for these cases remains under 10 seconds. Note that these results are based on a limited number of vehicle fleets in the system. The computation time could slightly increase if a significantly larger number of vehicles is considered, as a more coordinated decision-making process \cite{vaccaro2024achieving} would be required within the optimal power flow problem.

\section{Conclusions and Future Work}
\label{sec:con}

This paper explores the potential of integrating autonomous unmanned vehicles into power system operations. The problem fundamentally involves an integrated decision-making process of vehicles across both transportation and power networks. We first formulate an optimal routing problem that enables each autonomous vehicle to determine the optimal path and cost to each candidate destination. After that, we introduce a mixed-integer formulation to optimize autonomous vehicle integration in grid operations. To handle the nonlinearity, we employ McCormick relaxation, which provides a tractable, exact reformulation and facilitates real-time computations of the algorithm. Future work will include the rolling horizon formulation, which also accounts for vehicle battery state of charge, changing traffic and system conditions, dynamic pricing, as well as incentive mechanism design.

\vspace{1mm}

\bibliography{bibliography.bib}

% \vfill
% \begin{flushbottom} 
% \bibliography{bibliography.bib} 
% \end{flushbottom}

%\begin{comment}
%\begin{thebibliography}{1}

%\itemsep 2pt

%\bibitem{ANSI}
%emph{American National Standard
%For Electric Power Systems and Equipment- Voltage Ratings (60 Hertz)}, ANSI C84.1-2011, American National Standards Institute (ANSI), Inc. 

%\end{thebibliography}
%\end{comment}

\bibliographystyle{IEEEtran}

\end{document}